\documentclass[11pt]{elsarticle}

\pdfoutput=1

\makeatletter
\def\ps@pprintTitle{%
  \let\@oddhead\@empty
  \let\@evenhead\@empty
  \let\@oddfoot\@empty
  \let\@evenfoot\@oddfoot
}
\makeatother

\usepackage{url}
\usepackage{breakurl}
\usepackage[breaklinks,
            colorlinks = true,
            linkcolor = blue,
            urlcolor  = blue,
            citecolor = blue,
            anchorcolor = blue]{hyperref}

\usepackage{lineno}

\usepackage{graphicx}
\usepackage[margin=1.25in]{geometry}
\usepackage[usenames,dvipsnames]{color}

\newcommand\acp{\begin{center}
\rule[-0.2in]{\hsize}{0.01in}\\\rule{\hsize}{0.01in}\\
\vskip 0.1in The African Conference on Fundamental and Applied Physics\\ 
An Activity of the African School of Physics
    \vskip 0.05in
    {\it Second Edition, ACP2021, March 7--11, 2022 --- Virtual Event}\\
\rule{\hsize}{0.01in}\\\rule[+0.2in]{\hsize}{0.01in} \\
\end{center}}

\usepackage[firstpage=true]{background}
\backgroundsetup{contents={\parbox{6.5in}{\acp}}, scale=1,placement=top,opacity=1,color=black,position={3.25in,1.2in}}

\usepackage{fancyhdr}
\fancypagestyle{plain}{%
  \fancyhf{}%
  \fancyhead[C]{}
  \fancyfoot[C]{\thepage}
}

\fancypagestyle{empty}{%
  \fancyhf{}%
  \fancyhead[C]{{\it ACP2021, March 7--11, 2022 --- Virtual Edition}}
  \fancyfoot[C]{\thepage}
}
\begin{document}

\begin{frontmatter}


\title{Activity Report of the Second African Conference on Fundamental and Applied Physics, ACP2021}

\author[add1]{K\'et\'evi A. Assamagan\corref{cor1}}
\ead{ketevi@bnl.gov}
\author[add2]{Obinna Abah}
\author[add3]{Amare Abebe}
\author[add4]{Stephen Avery}
\author[add1]{Diallo Boye}
\author[add5]{Arame Boye-Faye}
\author[add6]{Kenneth Cecire}
\author[add7]{Mohamed Chabab}
\author[add8]{Samuel Chigome}
\author[add9]{Simon Connell}
\author[add10]{Mary Chantal Cyulinyana}
\author[add11]{Mark Macrae Dalton}
\author[add12]{Christine Darve}
\author[add7]{Lalla Btissam Drissi}
\author[add7]{Farida Fassi}
\author[add13]{Ulrich Goelach}
\author[add14]{Mohamed Gouighri}
\author[add15]{Paul Gueye}
\author[add16]{Sonia Haddad}
\author[add17]{Bjorn von der Heyden}
\author[add5]{Oumar Ka}
\author[add18]{Gihan Kamel}
\author[add19]{St\'ephane Kenmoe}
\author[add20]{Diouma Kobor}
\author[add21]{Tjaart Kr\"uger}
\author[add7]{Mounia Laassiri}
\author[add22]{Lerothodi Leeuw}
\author[add23]{Maria Moreno Llacer}
\author[add24]{Benard Mulilo}
\author[add25]{Robinson Musembi}
\author[add1]{Chilufya Mwewa}
\author[add26]{Clara Nellist}
\author[add27]{Lawrence Norris}
\author[add9]{Marie Cl\'ementine  Nibamureke}
\author[add28]{Mirjana Povi\'c}
\author[add29]{Horst Severini}
\author[add30]{Marco Silari}
\author[add31]{Bertrand Tchanche Fankam}
\author[add32]{Iyabo Usman}
\author[add33]{Esmeralda Yitamben}
\author[add34]{Jae Yu}

\cortext[cor1]{Corresponding Author}

\address[add1]{Brookhaven National Laboratory, USA}
\address[add2]{Newcastle University, UK}
\address[add3]{North-West University, South Africa}
\address[add4]{University of Pennsylvania, USA}
\address[add5]{Cheikh Anta Diop University, Senegal}
\address[add6]{University of Notre Dame, USA}
\address[add7]{Mohamed V University, Morocco}
\address[add8]{Botswana Institute for Technology Research and Innovation, Botswana}
\address[add9]{University of Johannesburg, South Africa}
\address[add10]{University of Rwanda, Rwanda}
\address[add11]{Thomas Jefferson National Accelerator Facility, USA}
\address[add12]{European Spallation Source, Sweden}
\address[add13]{Institut Pluridisciplinaire Hubert Curien CNRS, Strasbourg, France}
\address[add14]{Ibn-Tofail University, Morocco}
\address[add15]{Michigan State University, USA}
\address[add16]{University of Tunis El Manar, Tunisia}
\address[add17]{Stellenbosch University, South Africa}
\address[add18]{SESAME Jordan and Helwan University Egypt}
\address[add19]{Universitaet Duisburg-Essen, Germany}
\address[add20]{Universit\'e Assane Seck de Ziguinchor, Senegal}
\address[add21]{University of Pretoria, South Africa}
\address[add22]{University of the Western Cape, South Africa}
\address[add23]{IFIC institute (CSIC-UV), University of Valencia, Spain}
\address[add24]{University of Zambia, Zambia}
\address[add25]{University of Nairobi, Kenya}
\address[add26]{Radboud University and Nikhef, The Netherlands}
\address[add27]{African Light Source Foundation, USA}
\address[add28]{Ethiopian Space Science and Technology Institute, Ethiopia}
\address[add29]{University of Oklahoma, USA}
\address[add30]{CERN, Switzerland}
\address[add31]{Universit\'e Alioune Diop, Senegal}
\address[add32]{University of the Witwatwersrand, South Africa}
\address[add33]{Merck Group, Merck KGaA, Germany}
\address[add34]{University of Texas Arlington, USA}

\begin{abstract}
\noindent 
The African School of Fundamental Physics and Applications, also known as the African School of Physics (ASP), was initiated in 2010, as a three-week biennial event, to offer additional training in fundamental and applied physics to African students with a minimum of three-year university education. Since its inception, ASP has grown to be much more than a school. ASP has become a series of activities and events with directed ethos towards physics as an engine for development in Africa. One such activity of ASP is the African Conference on Fundamental and Applied Physics (ACP). The first edition of ACP took place during the 2018 edition of ASP at the University of Namibia in Windhoek. In this paper, we report on the second edition of ACP, organized on March 7--11, 2022, as a virtual event. 
\end{abstract}

\begin{keyword}
The African School of Physics \sep ASP \sep the African Conference on Fundamental and Applied Physics \sep ACP
\end{keyword}

\end{frontmatter}

%



\section{Introduction}
\label{sec:intro}
The  African School of Physics has been organized biennially in different African countries since 2010. It started as a three-week summer program in particle and (particle)astrophysics and related applications to complement education and research experiences of African students. The first edition of ASP was organized in South Africa and subsequent editions took place in Ghana, Senegal, Rwanda and Namibia in 2012, 2014, 2016 and 2018 respectively. The 2020 edition, originally planned in Morocco, was cancelled because of COVID-19 and reformatted as a virtual two-week school in 2021. Interest in ASP has increased from over a hundred applications in the 2010 edition to over several hundred by 2020. Detailed reports on ASP are presented in Refs.~\cite{ASP2021-reports, ASP, ASP-reports}. From 2016, the ASP program was augmented with a workshop for high school teachers and a physics outreach for high school pupils in the host country. A structured mentorship program was also added to provide further support to African students beyond the 2-3 weeks in-person events. In 2020, during the COVID-19 pandemic, the ASP program was further augmented with an online lecture series that run weekly or bi-weekly~\cite{ASP2021-reports}; ASP alumni also self-organized to study the COVID-19 pandemic in the home countries~\cite{ASP-COVID}.
During the 2016 edition in Rwanda, an extension of ASP to include a conference was discussed. This conference, called the African Conference on Fundamental and Applied Physics (ACP) would be of a one-week duration and run in parallel to other ASP activities within the  three-week engagements. The addition of ACP was needed to address growing interests by extending participation to the broader African and international communities. ACP allows for extended participation and scientific engagements, with the added benefit of increased networking among participants. After the initial discussions on ACP in 2016, the first edition (ACP2018) was integrated into the ASP program of 2018 (ASP2018) in Namibia~\cite{asp2018}. In addition to students, high school teachers, pupils and lecturers that participated in ASP2018, about sixty extra participants attended ACP2018. The second edition was planned to be integrated into the activities of the ASP2020 in Morocco, but that did not happen because of the COVID-19 pandemic. As mentioned above, school activities of ASP2020 were organized in 2021---the three-week event was shorten to two weeks in order to not have too long online engagements. The conference part of the activities, which was renamed ACP2021, was set in December 2021. However, the outbreak of Omicron forced the postponement of ACP2021 and was ultimately organized as a virtual event on March 7--11, 2022.
In this article, we report on the activities of ACP2021 in Section~\ref{sec:acp2021}. In Section~\ref{sec:conc}, we offer concluding remarks and an outlook.

\section{The Second African Conference on Fundamental and Applied Physics (ACP2021)}
\label{sec:acp2021}
\begin{figure}[!htbp]
\begin{center}
\includegraphics[width=\textwidth]{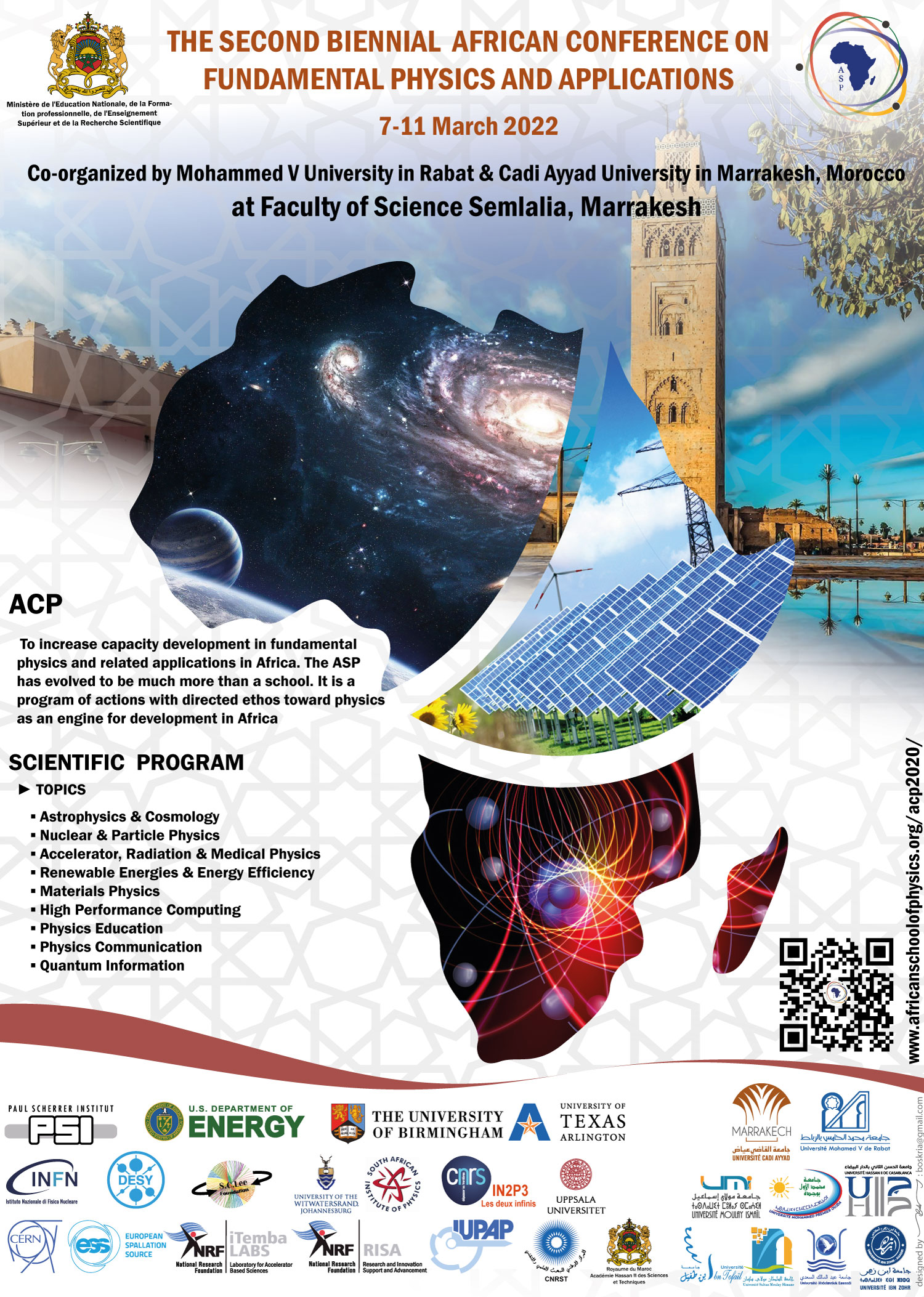}
\end{center}
\caption{Poster of the second African Conference on Fundamental and Applied Physics, ACP2021, co-organized with Cadi Ayyad University and Mohamed V University in Morocco.}
\label{fig:poster}
\end{figure}

As mentioned in Section~\ref{sec:intro}, the three-week event, planned for the sixth African School of Physics in Morocco, was reconfigured into two online events:
\begin{itemize}
  \item A two-week online school in July 2021~\cite{ASP-reports};
  \item A one-week online conference in March 2022, ACP2021~\cite{acp2021}.
\end{itemize}
Since ACP2021 was a part of sixth edition of ASP, hosted by Morocco, it was organized with Cadi Ayyad University~\cite{marrakesh} in Marrakesh and Mohammed V University~\cite{rabat} in Rabat. The poster of ACP2021 is shown in Figure~\ref{fig:poster}.
Since the aim of ACP is to extend of reach of ASP with broader participation in fundamental fields and related applications, the scientific program at ACP2021 included the major physics areas of interest in Africa, as defined by the African Physical Society (AfPS)~\cite{AfPS}:
\begin{itemize}
   \item Particles and related applications: nuclear physics, particle physics, medical physics, (particle)astrophysics \& cosmology, fluid \& plasma physics, complex systems;
   \item Light sources and their applications: light sources, condensed matter \& materials physics, atomic \& molecular physics, optics \& photonics, earth science;
   \item Cross-cutting fields: accelerator physics, computing, instrumentation \& detectors.
\end{itemize}
Topics in quantum computing \& quantum information and machine learning \& artificial intelligence were also on the agenda. Furthermore, ACP2021 included the fields of societal engagements, namely: topics related to physics education, community engagement, women in physics and early career physicists. Details on the scientific program are presented in Ref.~\cite{acp2021}, and further information about event is shown in Table~\ref{tab:table1}. 

\begin{figure}[!htb]
\begin{center}
\includegraphics[width=\textwidth]{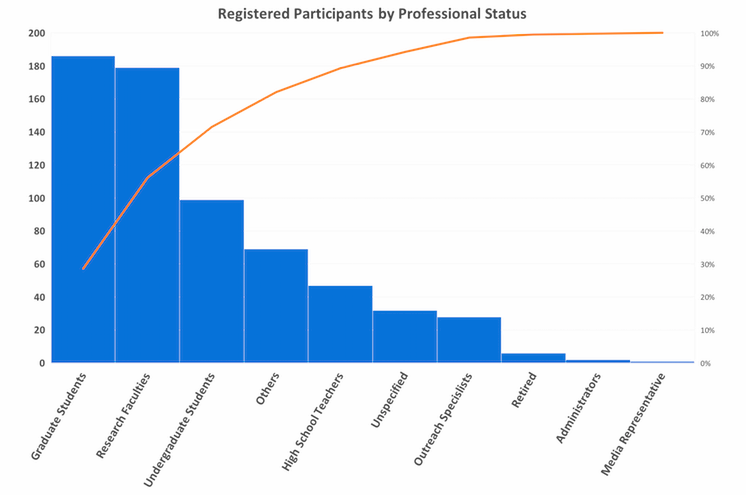}
\end{center}
\caption{Academic levels of registrants of ACP2021. Graduate students and research faculties were the largest participants. The “Unspecified” were folks that had registered previously for remote participation but did not change their registration type when the format of the event changed to fully virtual.}
\label{fig:levels}
\end{figure}

\begin{figure}[!htb]
\begin{center}
\includegraphics[width=\textwidth]{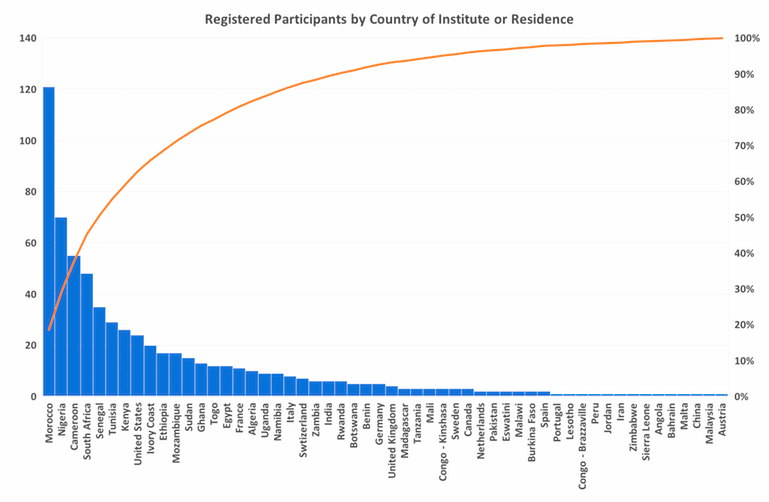}
\end{center}
\caption{Number of registered participants as a function of countries of residence or institutes of the participants.}
\label{fig:institutes}
\end{figure}

\begin{figure}[!htb]
\begin{center}
\includegraphics[width=\textwidth]{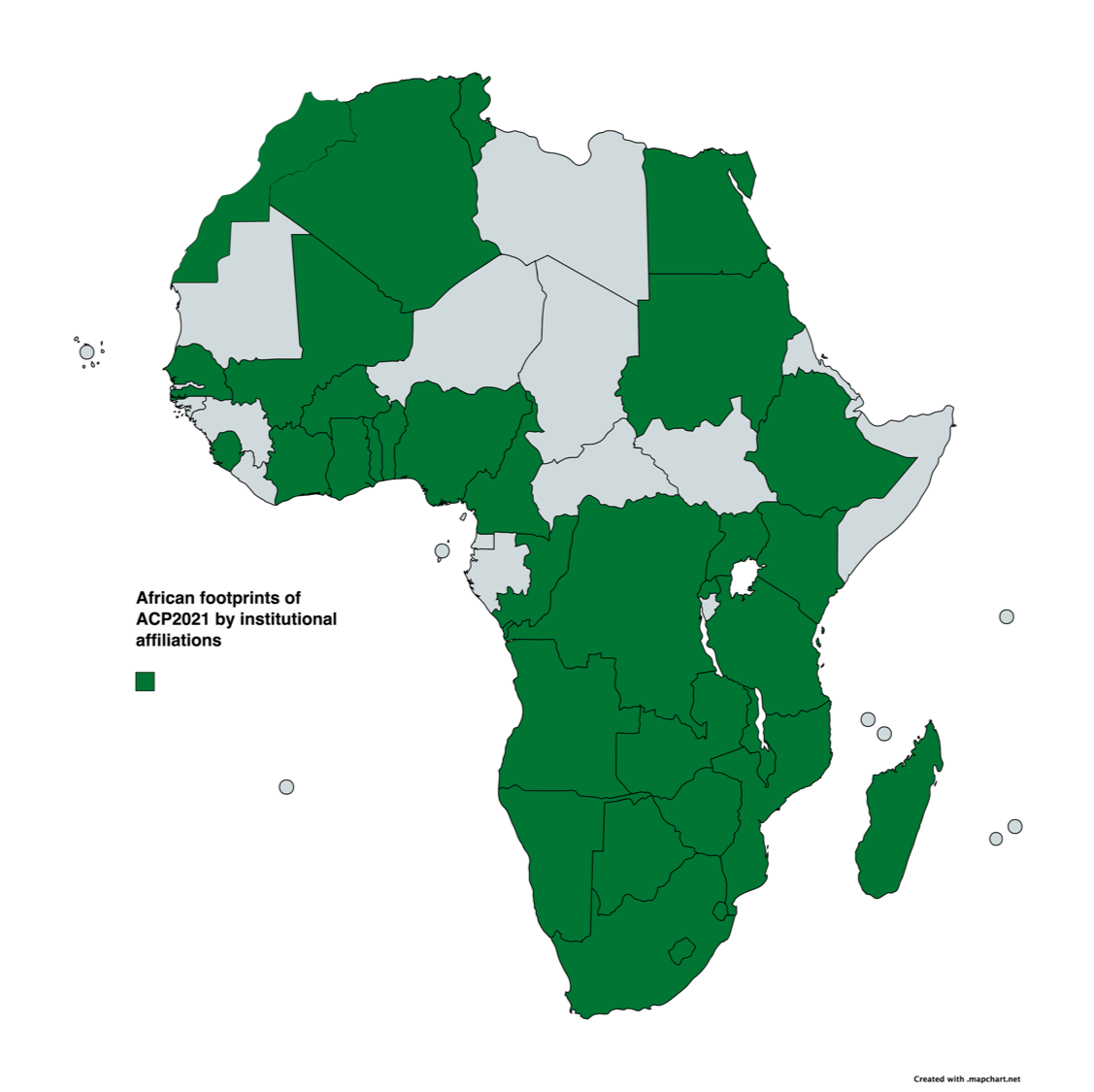}
\end{center}
\caption{The African countries where the institutes of the registered participants were located.}
\label{fig:reach}
\end{figure}

\begin{table}[!htbp]
\centering
\begin{tabular}{l|l}
\hline\hline
 \textbf{Event Information}   &  \textbf{Number of Engagements} \\ \hline
 Total registered participants & 649 \\
 Peak number of online connections & 191 \\
 Minimum connections in plenary sessions & 120 \\
 Participants from outside Africa & 86 \\
 Participants from Africa & 563 \\
 Number of countries represented & 54 \\
 Number of African countries represented & 33 \\ 
 \hline\hline
\end{tabular}
\caption{The organizers of ACP2021 made a remarkable effort to reach out to the broader African and internal communities. Level of participation increased over an order of magnitude compared to first ACP in 2018.}
\label{tab:table1}
\end{table}
The profiles of the registered participants, in terms of their professional standings, are shown in Figure~\ref{fig:levels}.
The registrations by country of residence or institutes are shown in Figure~\ref{fig:institutes}.
To see the reach of ACP2021, we show in Figure~\ref{fig:reach} the African countries where the institutes of the participants are located---thirty-three out of fifty-four African countries were represented at the second African conference on fundamental and applied physics.
The scientific program consisted of thirty-two invited talks in the aforementioned topics, one panel discussion on physics education and research in Africa, fifty contributed oral presentations in parallel sessions, thirty poster presentations in parallel sessions, one colloquium on ``Physics for sustainable development in Africa'', and three working sessions with three summary reports on the African Strategy for Fundamental and Applied Physics (ASFAP)~\cite{asfap}---as shown in the scientific program agenda~\cite{acp2021}. A selection committee made of the parallel session chairs and organizers followed the contributed oral and poster presentations, and the three best oral and three best poster presentations were identified; the committee reported its decisions in a dedicated session on the last day of the event. The winners of the contributed oral presentations had institutional affiliations in Cameroon, Morocco and South Africa, with topics in astrophysics \& cosmology, medical physics and nuclear physics, respectively. For poster presentations, the three winners were affiliated to institutes in Morocco (two) and South Africa, with topics in particle physics detector \& instrumentation, particle physics theory and materials physics, respectively. Each winner of the oral or poster presentations was entitled to a thousand Euros to support physics education or research activities. 

Online networking was arranged during breaks to increase interactions among participants. Upon request, some participants received a certificate of participation.

Invited speakers and panelists included representatives from the African Academy of Sciences~\cite{aas}, UNESCO~\cite{unesco}, IAEA~\cite{iaea}, University of Cambridge~\cite{cambridge}, Paul Scherrer Institute~\cite{psi}, DESY~\cite{desy}, Cheikh Anta Diop University~\cite{ucad}, ICTP~\cite{ICTP}, and the Democratic Republic of Congo Presidential Panel to the African Union and Investing In People ASBL~\cite{IIP}.

A few group pictures taken during ACP2021 are shown in the appendix.

\section{Conclusions}
\label{sec:conc}
The African Conference on Fundamental and Applied Physics is an event integrated into the activities of the African School of Physics since 2018. The second edition of ACP took place on March 7--11, 2022, as a virtual event and drew about six hundred and fifty registered participants, five hundred and sixty-three of whom came from thirty-three African countries. Compared to the first edition in 2018, which was an in-person event, the number of participants increased by an order of magnitude in the virtual second edition. This increase was due to two reasons: 1) the virtual arrangement eliminated the financial burden associated to international travels, thus allowed more participation; 2) the efforts of the organizing committee to advertise the event broadly through different network outlets. The scientific program included the physics research areas of interest to Africa as defined by AfPS and societal engagement topics that impact physics education and research in Africa. Graduate students and research faculties formed the largest fractions of the participants. The event included invited and contributed presentations, a panel discussion, a colloquium and poster sessions. The event also featured working and summary sessions on ASFAP. Interactions among participants were encouraged during breaks in organized online networking. 

The next edition of ASP will be hosted by South Africa; it will be arranged in two events: 1) a two-week physics school on November 28 -- December 9, 2022; 2) the third African Conference on Fundamental and Applied physics in 2023, ACP2022. The exact date of ACP2022 will be announced later. 

\section*{Acknowledgments}
The organizers would like to thank:
\begin{itemize}
\item the institutes that supported ACP2021, in-kind or financially, namely Cadi Ayyad University, Mohamed V University, the South African Institute of Physics, Brookhaven National Laboratory, CERN and the U.S. Department of Energy Office of Science. ACP2021 could not have happened, so succesfully, without their support;
\item the invited delegates in the plenary sessions and panel discussion;
\item all the people who contributed abstracts and made oral or poster presentations;
\item the ASFAP community for enriching working sessions;
\item the selection committee for best oral and poster presentations;
\item all the participants;
\item each member event organizing team; their diligence and dedication made ACP2021 a success.
\end{itemize}

\newpage
\section*{Appendix}
We show here a couple of pictures of the participants, pictures taken during the event.
\begin{figure}[!h]
\begin{center}
\includegraphics[width=\textwidth]{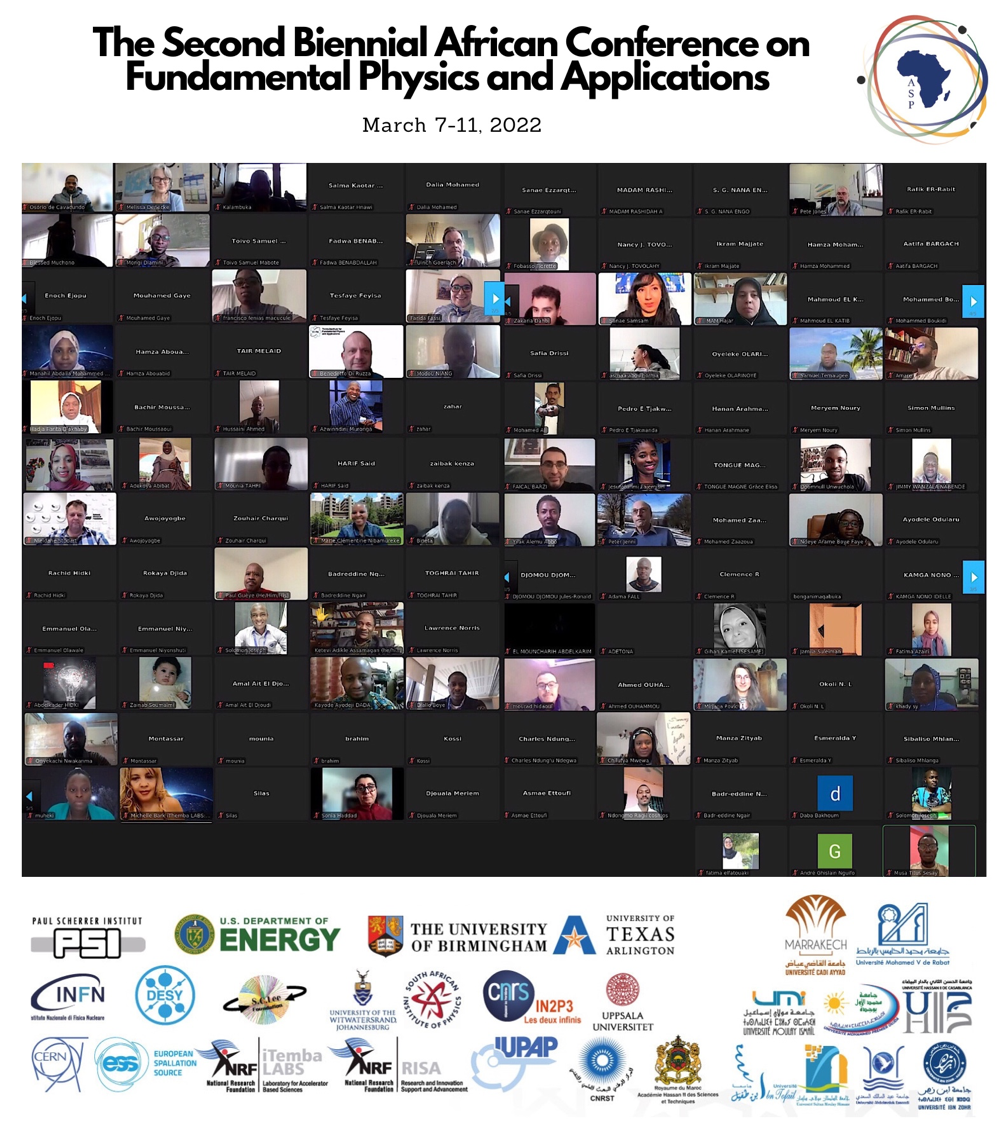}
\end{center}
\caption{Participants connected to the event on March 11, 2022}
\label{fig:march11}
\end{figure}

\begin{figure}[!p]
\begin{center}
\includegraphics[width=\textwidth]{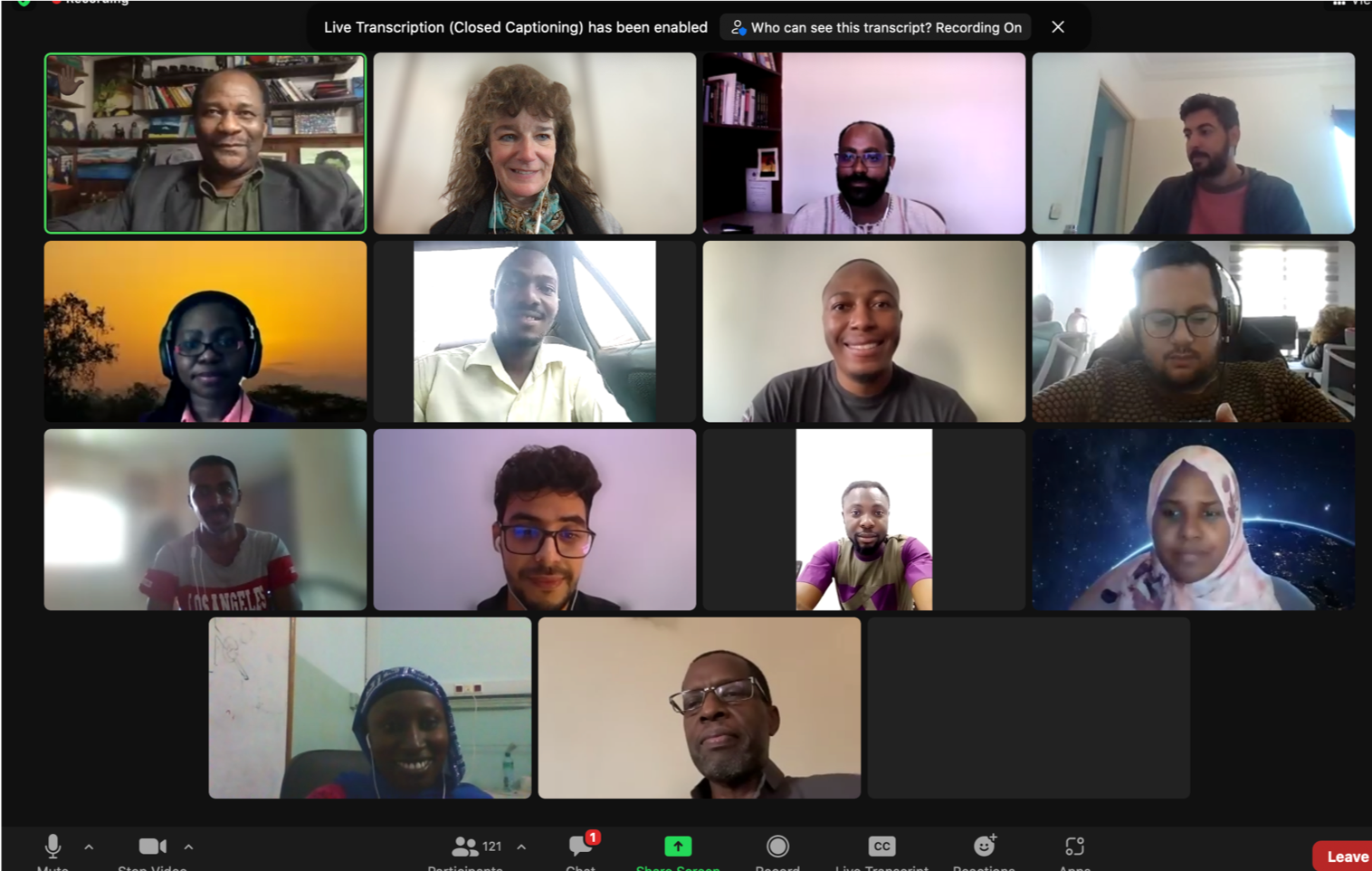}
\end{center}
\caption{Some of the winners of the oral and poster presentations, with members of the selection committee.}
\label{fig:winners}
\end{figure}

\newpage

\bibliographystyle{elsarticle-num}
\bibliography{myreferences} 

\end{document}